\documentclass[10pt, final, conference, letterpaper]{IEEEtran}
\usepackage[left=.6in,right=.6in,top=.719in,bottom=.99in]{geometry}

\pdfoutput=1

\usepackage{amsmath,amssymb,dsfont,stfloats,color,url,bbm}
\usepackage[pdftex]{graphicx}
\usepackage[caption=false,font=footnotesize]{subfig}
\usepackage{mathtools}

\DeclareGraphicsExtensions{.eps,.pdf,.png,.jpg,.gif,.jpeg,.pstex}

\makeatletter
\def\Hline{%
\noalign{\ifnum0=`}\fi\hrule \@height 2pt \futurelet
\reserved@a\@xhline}
\makeatother

\usepackage[noadjust]{cite}
\usepackage{url}

\newfont{\bbb}{msbm10 scaled 700}

\newfont{\bb}{msbm10 scaled 1100}
\newcommand{\CC}{\mbox{\bb C}}

\newcommand{\EE}{\mbox{\bb E}}

\newcommand{\HH}{\mbox{\bb H}}

\newcommand{\yy}{\mathbbm{y}}

\newcommand{\zz}{\mathbbm{z}}
\newcommand{\sss}{\mathbbm{s}}

\newcommand{\hh}{\mathbbm{h}}

\newcommand{\vvv}{\mathbbm{v}}

\newcommand{\av}{{\bf a}}

\newcommand{\hv}{{\bf h}}

\newcommand{\wv}{{\bf w}}
\newcommand{\vv}{{\bf v}}

\newcommand{\yv}{{\bf y}}
\newcommand{\zv}{{\bf z}}
\newcommand{\zerov}{{\bf 0}}

\newcommand{\Dm}{{\bf D}}

\newcommand{\Fm}{{\bf F}}
\newcommand{\Gm}{{\bf G}}

\newcommand{\Id}{{\bf I}}

\newcommand{\Ym}{{\bf Y}}
\newcommand{\Zm}{{\bf Z}}

\newcommand{\Cc}{{\cal C}}

\newcommand{\Ec}{{\cal E}}

\newcommand{\Gc}{{\cal G}}

\newcommand{\Lc}{{\cal L}}

\newcommand{\Nc}{{\cal N}}

\newcommand{\Pc}{{\cal P}}

\newcommand{\Sc}{{\cal S}}
\newcommand{\Tc}{{\cal T}}
\newcommand{\Uc}{{\cal U}}

\newcommand{\Vc}{{\cal V}}

\newcommand{\nuv}{\hbox{\boldmath$\nu$}}

\newcommand{\phiv}{\hbox{\boldmath$\phi$}}

\newcommand{\Gammam}{\hbox{\boldmath$\Gamma$}}

\newcommand{\Sigmam}{\hbox{\boldmath$\Sigma$}}

\newcommand{\diag}{{\hbox{diag}}}

\renewcommand{\arg}{{\hbox{arg}}}

\newcommand{\eqdef}{\stackrel{\Delta}{=}}

\newcommand{\herm}{{\sf H}}

\newcommand{\SINR}{{\sf SINR}}
\newcommand{\SNR}{{\sf SNR}}

\usepackage{stmaryrd} 

\renewcommand{\arg}{{\rm arg}}

\begin{document}

\setlength{\abovedisplayskip}{1pt}
\setlength{\belowdisplayskip}{1pt}
\setlength{\abovedisplayshortskip}{1pt}
\setlength{\belowdisplayshortskip}{1pt}

 \onecolumn
 
 This work has been submitted to the IEEE for possible publication. Copyright may be transferred without notice, after which this version may no longer be accessible.
 
 \newpage
 \twocolumn 

\title{Overloaded Pilot Assignment with Pilot Decontamination for Cell-Free Systems}

\author{\IEEEauthorblockN{Authors}
\IEEEauthorblockA{\IEEEauthorrefmark{1}KDDI Research Inc., Japan\\
\IEEEauthorrefmark{2}Technical University of Berlin, Germany\\
Emails: }}

\author{\IEEEauthorblockN{
		Noboru Osawa\IEEEauthorrefmark{1}, Fabian G\"ottsch\IEEEauthorrefmark{2}, Issei Kanno\IEEEauthorrefmark{1}, Takeo Ohseki\IEEEauthorrefmark{1}, \\ Yoshiaki Amano\IEEEauthorrefmark{1},   Kosuke Yamazaki\IEEEauthorrefmark{1}, Giuseppe Caire\IEEEauthorrefmark{2}}
	\IEEEauthorblockA{\IEEEauthorrefmark{1}KDDI Research Inc., Saitama, Japan\\
	\IEEEauthorrefmark{2}Technical University of Berlin, Germany\\
		Emails: \{nb-oosawa, is-kanno, ohseki, yo-amano,  ko-yamazaki\}@kddi.com, \{fabian.goettsch, caire\}@tu-berlin.de}}

\maketitle

\begin{abstract}

The pilot contamination in cell-free massive multiple-input-multiple-output (CF-mMIMO) must be addressed for accommodating a large number of users.
In previous works, we have investigated a decontamination method called subspace projection (SP). The SP separates interference from co-pilot users by using the orthogonality of the principal components of the users' channel subspaces.
Non-overloaded pilot assignment (PA), where each radio unit (RU) does not assign the same pilot to different users, limits the spectral efficiency (SE) of the system, since SP channel estimation is able to deal with co-pilot users that have nearly orthogonal subspaces.
Motivated by this limitation, this paper introduces overloaded PA methods adjusted for the decontamination in order to improve the sum SE of CF systems.
Numerical simulations show that the overloaded PA methods give higher SE than that of non-overloaded PA at a high user density scenario.

\end{abstract}


\section{Introduction} 

A large number of works in wireless communication theory is dedicated to the joint processing of spatially distributed antennas. This idea can be traced back to the work of Wyner \cite{wyner1994shannon}, and has been “re-marketed” several times under different names with slight nuances, such as coordinate multipoint (CoMP), cloud radio access network (CRAN), and currently, it is promoted as cell-free massive MIMO (CF-mMIMO).
As the name implies, one of the purposes of CF-mMIMO is eliminating cell boundaries and preventing user equipment's (UE's) performance degradation depending on their geographical positions.

The channel state information (CSI) acquisition with pilot signals at the infrastructure antenna side is important for taking advantage of CF-mMIMO's spatial multiplexing gain.
In this paper, we focus on uplink (UL) channel estimation in a time division duplex (TDD) system. Thanks to TDD operations and channel reciprocity, these estimates can also be used for the downlink precoding \cite{marzetta2010noncooperative}.
Generally, the number of orthogonal pilots is limited to reduce overhead and keep the training phase within coherence time.
Therefore, pilot reuse is inevitable to increase the number of UEs to be accommodated, and it induces pilot contamination.
In addition, CF-mMIMO system has no clear boundaries, so it is hard to apply a cell-based pilot reuse restriction.


As one of the pilot decontamination methods, in \cite{goettsch2021impact,goettsch_WCNC2022}, we have investigated a subspace projection (SP) based channel estimation, which uses receiver side discrete Fourier transform (DFT) processing.
The channel subspace is spanned by dominant components of channel covariance matrices.
Assuming uniform linear arrays (ULAs) or uniform planar arrays (UPAs), the channel covariance matrix is Toeplitz (for ULA) or Block-Toeplitz (for UPA), which both are approximately diagonalized by discrete Fourier transforms (DFT) on the columns and rows
\cite{adhikary2013joint}.
This means that the dominant channel subspaces are approximately spanned by subsets of the DFT columns.
If the antennas have correlation due to a limited scattering angular spread, each radio unit (RU) and UE link has an individual channel subspace.
Therefore, DFT-based projection onto the target UE's subspace can reduce interference from co-pilot UEs if the subspaces of the co-pilot UEs are nearly orthogonal, which we call decontamination.
We have shown that the effect of the pilot contamination can be reduced significantly with the SP-based decontamination, yielding a system performance close to the case of ideal CSI, where each RU has perfect knowledge of the channel vectors of its associated UEs \cite{goettsch2021impact}.

The system model of CF in our previous works \cite{goettsch2021impact, goettsch_SPbased_arxiv} is similar to  Bj\"{o}rnson and Sanguinetti's scalable CF system \cite{Bjornson2020Scalable}, where each user-centric cluster is formed by a finite number of RUs.
In  \cite{goettsch2021impact, goettsch_SPbased_arxiv}, we have not allowed the RUs to assign the same pilot to different users, which we call non-overloaded pilot assignment (PA).
However, the SP enhances RUs to estimate channels of co-pilot UEs under overloaded PA, where RUs can assign the same pilot to multiple UEs.
Adopting the overloaded PA leads to an increased number of associated RUs per UE, which can increase the spatial multiplexing gain per UE and improve spectral efficiency (SE).

Most of related works, which aimed to address contamination by strategic PA methods, implicitly adopted overloaded PA
because pilot selection is significantly constrained by the cluster formation, when the non-overloaded PA is adopted.
In \cite{Greedy}, a greedy PA algorithm was introduced as an early study in the CF literature.
In \cite{Hungarian}, a PA scheme aiming at user throughput maximization is investigated, where the maximization problem is solved by using an iterative scheme based on the Hungarian algorithm. 
In \cite{Chen_Structured_CF2021}, user-group PA was proposed which assigns the same pilot to UEs who share the fewest number of serving RUs.
Furthermore, graphic framework based pilot assignment schemes are investigated in \cite{Hmida_Graph_2020,Zeng_WGF_2021}.
Since these works do not adopt the SP-based decontamination, combining these overloaded PA methods and our decontamination method has potential to achieve better performance.

As mentioned above, to the best of our knowledge, combination of the SP decontamination and overloaded PA has not yet been investigated.
Motivated by this, we propose two types of overloaded PA methods in this paper.
We call these methods rough overloaded PA (R-OPA) and subspace information aided overloaded PA (SIA-OPA).
The R-OPA forms clusters in advance without considering the contamination, and then assigns the pilots.
We propose a specific implementation of R-OPA by adjusting a graphic framework based PA studied by Zeng et al. \cite{Zeng_WGF_2021} to our SP based decontamination.
The SIA-OPA has a restriction in the cluster formation phase that it adds an RU to a cluster of a UE only when the subspaces of co-pilot UEs are orthogonal.


The contributions of this work are summarized as follows.
\begin{itemize}
	\item We propose R-OPA and SIA-OPA for CF systems. In addition, a graphic framework-based PA is designed as a specific implementation of R-OPA.
    \item We discuss the complexity of the proposed PA.
    Note that R-OPA with the graphic framework-based PA algorithm requires network-wide information exchange, whereas SIA-OPA does not.
	\item We compare the sum SE and outage performance of the non-overloaded PA, R-OPA and SIA-OPA to reveal a region of parameters where the performance is best.
\end{itemize}

\section{System model}
We consider a CF-mMIMO system with $L$ RUs, each with $M$ antennas, and $K$ UEs. Both RUs and UEs are distributed on a squared region on the 2-dimensional plane.
The set $\Cc_k \subseteq [L] = \{1,2,\dots, L\}$ denotes the cluster of RUs that serve UE $k$, and $\Uc_\ell \subseteq [K]$ denotes the set of UEs served by RU $\ell$, where the set of integers from 1 to $n$ is denoted by $[n]$.
The RU-UE associations are described by a bipartite graph $\Gc$ whose vertices are the RUs and UEs, respectively. The set of edges accounting for associated RU-UE pairs is denoted by $\Ec$, i.e., $\Gc = \Gc([L], [K], \Ec)$.
We also define $\Uc(\Cc_k) = \bigcup_{\ell \in \Cc_k} \Uc_\ell$ as the set of UEs served by at least one RU in $\Cc_k$.

In UL transmissions, the UEs transmit with the same energy per symbol $E_{\rm s}$, and we define the system parameter
\begin{equation} 
	{\rm SNR} = \frac{E_{\rm s}}{N_0} ,
\end{equation}
where $N_0$ denotes noise power spectral density.
The large-scale-fading-coefficient (LSFC) between RU $\ell$ and UE $k$ is denoted by $\beta_{\ell,k}$ and includes pathloss, blocking effects and shadowing, respectively.
By assuming isotropic antennas, the maximum beamforming gain averaged over the small scale fading is $M$, therefore the maximum SNR at the receiver of RU $\ell$ from UE $k$ is $\beta_{\ell,k}M{\rm SNR}$.

We consider orthogonal frequency-division multiplexing (OFDM) modulation and channels following the standard block-fading model \cite{marzetta2010noncooperative,9336188,9064545}, such that they are random but constant over coherence blocks of $T$ signal dimensions in the time-frequency domain. 
The described methods are formulated for one resource block (RB), so the RB index is omitted for simplicity.
We define channel matrices and its elements as follows. $\HH \in \CC^{LM \times K}$ denotes the overall channel matrix between all $LM$ RU antennas and all $K$ UE antennas on a given RB. Next, $\hh_k \in \CC^{LM \times 1}$ denotes $k$-th column of $\HH \in \CC^{LM \times K}$. 
In addition, $\hv_{\ell,k} \in \CC^{M \times 1}$ denotes the channel vector between RU $\ell$ and UE $k$. 
Finally, $\HH(\Cc_k) \in \CC^{LM \times K}$ denotes the partial cluster-centric matrix for a cluster $\Cc_k$, whose $M \times 1$ blocks of RU-UE pairs $(\ell,k) \in \Ec$ are equal to $\hv_{\ell,k}$, and equal to $\zerov$ (the identically zero vector) otherwise.

For ease of analysis, we assume that the individual channels between RUs and UEs follow the single ring local scattering model \cite{adhikary2013joint} where the covariance matrix is perfectly diagonalized by DFT.
Then $\hv_{\ell,k}$ is given by
\begin{equation} 
	\hv_{\ell,k} = \sqrt{\frac{\beta_{\ell,k} M}{|\Sc_{\ell,k}|}}  \Fm_{\ell,k} \nuv_{\ell, k},
\end{equation}
where $\Sc_{\ell,k} \subseteq \{0,\ldots, M-1\}$ and $\nuv_{\ell,k}$ are the angular support set according to \cite{adhikary2013joint}, and an $|\Sc_{\ell,k}| \times 1$ i.i.d. Gaussian vector with components $\sim \Cc\Nc(0,1)$, respectively.
The set $\Sc_{\ell,k}$ is constructed by the angular support $\Theta_{\ell,k} = [\theta_{\ell,k} - \Delta/2, \theta_{\ell,k} + \Delta/2]$ centered at angle $\theta_{\ell,k}$ of the LOS between RU $\ell$ and UE $k$, with angular spread $\Delta$.
We let $\Fm$ denote the $M \times M$ unitary DFT matrix with $(m,n)$-elements $f_{m,n} = \frac{e^{-j\frac{2\pi}{M} mn}}{\sqrt{M}}$, and using a Matlab-like notation, $\Fm_{\ell,k} \eqdef \Fm(: , \Sc_{\ell,k})$ denotes the tall unitary matrix obtained by selecting the columns of $\Fm$ corresponding to the index set $\Sc_{\ell,k}$.

\subsection{UL data transmission}
Let us note that the UEs transmit with same power in the UL. 
In general cellular network systems, UL transmit power is controlled per UE mainly for 1) countering the near-far problem for the cell edge UEs, or 2) saving power for UEs close to an RU. 
We believe that the UL power control for the ``near-far problem'' hardly makes sense for CF-mMIMO systems, since the CF system itself tries to solve the near-far problem by making UEs connect to multiple RUs.
UL power control for energy saving or rate maximization still can be considered, however, it is very challenging to implement a centralized process for optimization in a distributed system or to find a decentralized way of UL power control, which will be future work.

According to the results in \cite{goettsch2021impact}, we choose the local LMMSE with cluster-level combining for this work, since it showed high performance with scalability, compared to other methods such as cluster-level zero-forcing and local MRC.

The received $LM \times 1$ symbol vector at the $LM$ RU antennas in UL is given by
\begin{equation} 
	\yy^{\rm ul} = \sqrt{\SNR} \; \HH \sss^{\rm ul}   + \zz^{\rm ul}, \label{ULchannel}
\end{equation}
where $\sss^{\rm ul} \in \CC^{K \times 1}$ is the vector of information symbols transmitted by the UEs and $\zz^{\rm ul}$ is an i.i.d. noise vector with components $\sim \Cc\Nc(0,1)$.
We define the received symbols vector at RU $\ell$ as $\yy^{\rm ul}_{\ell} \in \CC^{M \times 1}$, and the receiver unit norm vector for UE $k$ as $\vvv_k \in \CC^{LM \times 1}$ formed by $M \times 1$ blocks $\vv_{\ell,k} : \ell = 1, \ldots, L$, such that $\vv_{\ell,k} = \zerov$ if $(\ell,k) \notin \Ec$. 
%
Each RU $\ell \in \Cc_k$ computes the local LMMSE receiving vector $\vv_{\ell,k}$ for the UEs $k \in \Uc_\ell$; given by
\begin{equation} 
	\vv_{\ell,k} = \left ( \sigma_\ell^2 \Id + \SNR \sum_{j \in \Uc_\ell} \hv_{\ell,j} \hv_{\ell,j}^\herm \right )^{-1} \hv_{\ell,k},  \label{eq:lmmse}
\end{equation}
where $\sigma_\ell^2$ denotes the approximated variance of noise and external interference \cite{goettsch_SPbased_arxiv}, given by
\begin{equation} 
	\sigma^2_\ell =  1 + \SNR \sum_{j \neq \Uc_\ell}  \beta_{\ell,j}. \label{sigmaell} 
\end{equation}

RU $\ell$ computes the local observation $r^{\rm ul}_{\ell,k} = \vv_{\ell,k}^{\herm}\yv^{\rm ul}_{\ell}$ for each $k \in \Uc_\ell$ and
sends the observations to a centralized unit for the cluster-level combining. Then the centralized unit computes the cluster-level combined observation as
\begin{equation} 
	r^{\rm ul}_k = \sum_{\ell \in \Cc_k} w_{\ell,k}^{*} r^{\rm ul}_{\ell,k}.   \label{combining_perfect_csi}
\end{equation}
The optimized combining weights vector $\wv_k = \{w_{\ell,k} : \ell \in \Cc_k\}$, which maximizes the SINR \cite{goettsch2021impact}, is given by 
\begin{eqnarray}
	\wv_k &=& \Gammam_k^{-1} \av_k \ \in \CC^{|\Cc_k| \times 1} ,\\
	\av_k &=& \{ g_{\ell, k,k} : \ell \in \Cc_k\} ,
\\
	 g_{\ell,k,j} &=& \vv_{\ell,k}^\herm \hv_{\ell,j} ,\\
	\Gammam_k &=& \Dm_k  + \SNR \; \Gm_k \Gm^\herm_k \ \in \CC^{|\Cc_k| \times |\Cc_k|},\\
	\Dm_k &=& \diag \left \{ \sigma_\ell^2 \|\vv_{\ell,k}\|^2 : \ell \in \Cc_k \right \},
\end{eqnarray} 
where the matrix $\Gm_k$ of dimension $|\Cc_k| \times ( | \Uc(\Cc_k)| - 1)$ contains elements $g_{\ell, k, j}$ in position corresponding to RU $\ell$ and UE $j$ (after a suitable index reordering) if $(\ell,j) \in \Ec$, and zero elsewhere.
The overall receiving vector $\vvv_k$ is formed by stacking the vectors $w_{\ell,k}\vv_{\ell,k}$, i.e., $\vvv_k = [w_{1,k}\vv_{1,k}^\mathrm{T}, w_{2,k}\vv_{2,k}^\mathrm{T}, ..., w_{L,k}\vv_{L,k}^\mathrm{T} ]^\mathrm{T}$.



The resulting SINR for UE $k$'s UL symbol is given by
\begin{eqnarray} 
	\SINR^{\rm ul}_k 
& = & \frac{  |\vvv_k^\herm \hh_k|^2 }{ \SNR^{-1}  + \sum_{j \neq k} |\vvv_k^\herm \hh_j |^2 }.  \label{UL-SINR-unitnorm}
\end{eqnarray}
We use the optimistic ergodic achievable rate $R_k^{\rm ul}$ for performance evaluation, which is given by
\begin{equation} 
R_k^{\rm ul} = \EE \left[ \log( 1+\SINR^{\rm ul}_k ) \right],
\end{equation}
where the expectation is with respect to the small scale fading.
Then, the UL spectral efficiency (SE) is calculated as
\begin{equation} 
{\rm SE}_k^{\rm ul} = (1-\tau_p/T)R_k^{\rm ul},
\label{eq:SE}
\end{equation}
where $T$ is the dimension of an RB and $\tau_p$ is the pilot dimension.

\subsection{UL channel estimation}
As a practical remark, we note that in 5GNR two types of UL pilots are specified, the demodulation reference signals (DMRS) and the sounding reference signals (SRS). In this work we assume that the instantaneous channel coefficients are estimated from orthogonal DMRS pilot sequences, and the subspace information is estimated by utilizing SRS pilots. \footnote{An estimation method of subspace information by utilizing SRS pilots is discussed in \cite{goettsch_SPbased_arxiv}. }
%

The DMRS pilot field received at RU $\ell$ is given by the $M \times \tau_p$ matrix  as
\begin{equation} 
\Ym_\ell^{\rm DMRS} = \sum_{i=1}^K \hv_{\ell,i} \phiv_{t_i}^\herm + \Zm_\ell^{\rm DMRS}, \label{Y_pilot} 
\end{equation}
where $\phiv_{t_i}$ denotes the DMRS pilot vector with index $t_i$ of dimension $\tau_p$ used by UE $i$ in the current RB, with total energy $\| \phiv_{t_i} \|^2 = \tau_p \SNR$.
For each UE $k \in \Uc_\ell$, RU $\ell$ produces the {\em pilot matching} (PM) channel estimate
\begin{eqnarray} 
	\widehat{\hv}^{\rm pm}_{\ell,k} & = & \frac{1}{\tau_p \SNR} \Ym^{\rm DMRS}_\ell \phiv_{t_k} \nonumber \\ 
	& = & \hv_{\ell,k}  + \sum_{\substack{i : t_i = t_k \\ i\neq k}} \hv_{\ell,i}  + \widetilde{\zv}_{t_k,\ell} ,  \label{chest}
\end{eqnarray}
where $\widetilde{\zv}_{t_k,\ell}$ has i.i.d. with components $\Cc\Nc(0, \frac{1}{\tau_p\SNR})$. Notice that the presence of UEs $i \neq k$ using the same DMRS pilot $t_k$ yields pilot contamination. 

We consider the SP based decontamination scheme for which the projected channel estimate is given by the orthogonal projection of $\widehat{\hv}^{\rm pm}_{\ell,k}$ onto the subspace spanned by the columns of $\Fm_{\ell,k}$, i.e., 
\begin{align}
	\widehat{\hv}^{\rm sp}_{\ell,k} &= \Fm_{\ell,k}\Fm_{\ell,k}^\herm \widehat{\hv}^{\rm pm}_{\ell,k} \nonumber \\
	& = \hv_{\ell,k}  + \sum_{\substack{i : t_i = t_k \\ i\neq k}} \Fm_{\ell,k}\Fm_{\ell,k}^\herm \hv_{\ell,i}  + \Fm_{\ell,k} \Fm_{\ell,k}^\herm \widetilde{\zv}_{t_k,\ell}  .
	\label{chest_sp}
\end{align}
The second term of the last equation corresponds to pilot contamination after the SP, which is a Gaussian vector with zero mean and its covariance matrix can be written as
\begin{equation}
	\Sigmam_{\ell,k}^{\rm co}  = \sum_{\substack{i : t_i = t_k \\ i\neq k}} \frac{\beta_{\ell,i} M}{|\Sc_{\ell,i}|}  \Fm_{\ell,k} \Fm_{\ell,k}^\herm \Fm_{\ell,i} \Fm^\herm_{\ell,i} \Fm_{\ell,k} \Fm^\herm_{\ell,k}. 
\end{equation}
When $\Fm_{\ell,k}$ and $\Fm_{\ell,i}$ are nearly mutually orthogonal, i.e. $\Fm_{\ell,k}^\herm \Fm_{\ell.i} \approx \zerov$,
the subspace projection is able to reduce the pilot contamination effect.

\section{Pilot assignment and cluster formation}


As mentioned in Sec. 1, the SP can enhance overloaded PA, and lead to improved SE compared to non-overloaded PA.
This section introduces the non-overloaded PA, R-OPA and SIA-OPA schemes with their cluster formation rules.

\subsection{Non-overloaded pilot assignment}
\subsubsection{Leader RU selection}

When a UE $k$ wishes to join the system, it selects its leading RU $\ell(k)$ as the one with the largest LSFC and free DMRS pilots.
This is formulated as $\ell(k) = \underset{\ell \in\Lc_{\rm f}}{\arg\max}\  \beta_{\ell,k}$, where $\Lc_{\rm f}$ denotes the set of RUs with free DMRS pilots.
The leader RU selection also needs to satisfy the condition
\begin{equation}
	\beta_{\ell,k} \geq \frac{\eta}{M \SNR} , \label{eq:snr_threshold}
\end{equation}
where  $\eta > 0$ is a suitable threshold determining how much above the noise floor the useful signal in the presence of maximum possible beamforming gain (equal to $M$) should be. 
If such RU is not available, then the UE is declared in outage.

\subsubsection{Pilot selection}
The RU chooses the pilot with least interference for UE $k$, i.e., UE $k$'s pilot $t_k$ is given by
\begin{eqnarray}
    t_k &=& \mathop{\arg\min}\limits_{i \in \Tc_\ell} \sum_{j \in \Pc_i} \beta_{\ell,j},
    \label{eq:pilot_choice}
\end{eqnarray}
where $\Tc_\ell$ and $\Pc_i$ denote the set of not assigned pilots at RU $\ell$ and the set of UEs with pilot index $i$, respectively. The set $\Pc_i$ is updated as $\Pc_i = \Pc_i \cup k$ after assignment of $t_k$.
The RU only knows the result of $\sum_{j \in \Pc_i} \beta_{\ell,j}$ as measured statistics.

\subsubsection{Cluster formation}
Suppose that UE $k$ finds its leader RU $\ell(k)$ and it is allocated the DMRS pilot with index $t_k \in [\tau_p]$.
Then, the cluster $\Cc_k$ is obtained sorting the RUs satisfying condition (\ref{eq:snr_threshold}) and having pilot $t_k$ still available in decreasing LSFC order, and adding them to the cluster until a maximum cluster size $Q$ is reached, where $Q$ is a design parameter imposed to limit the computational complexity. 
As a result, for all UEs $k$ not in outage, $1 \leq  |\Cc_k| \leq Q$ and for all the RUs $\ell \in \Cc_k$, the corresponding LSFC satisfies 
(\ref{eq:snr_threshold}). Furthermore, for all RUs $\ell$ we have $|\Uc_\ell| \leq \tau_p$.


\subsection{Subspace information aided overloaded PA}

We introduce a SIA-OPA approach where each RU assigns the same pilot only to UEs with orthogonal subspaces.
Let us assume that the RUs acquire subspace information of UEs within its coverage before the leader RU selection phase.

\subsubsection{Leader RU selection}
In the leader RU selection phase, UE $k$ firstly try to choose the RU with the largest LSFC as $\ell(k)$.
If the RU $\ell(k)$ has free pilots, same as non-overloaded PA, the RU selects a pilot based on eq. (\ref{eq:pilot_choice}).
On the other hand, when the RU has no free pilot, the RU tries to find
a pilot, which is assigned only to UEs with an orthogonal subspace with respect to UE $k$'s subspace.

\subsubsection{Pilot selection}
A contamination quantity per pilot index ${i}$ on the subspace of the channel between RU $\ell(k)$ and UE $k$ is formulated by
\begin{eqnarray}
    \zeta_{\ell,k,{i}} &\triangleq& \sum_{j \in (\Uc_\ell \cap \Pc_{{i}} \setminus k)} || \Fm_{\ell,k}\Fm_{\ell,k}^\herm\Fm_{\ell,j} ||_F \beta_{\ell, j}.
    \label{eq:zeta}
\end{eqnarray}
At first, if $\zeta_{\ell,k,{i}} = 0$ due to that pilot index ${i}$ is not assigned to any other UEs in the coverage of RU $\ell(k)$,  the pilot can be assigned to UE $k$ as with the non-overloaded PA.
In addition, even if there are other UEs with pilot index ${i}$, the metric $\zeta_{\ell,k,{i}}$ also becomes 0 if the subspaces are orthogonal as $\Fm_{\ell,k}^\herm \Fm_{\ell,j}$ results in a zero matrix.
Therefore, with SIA-OPA, the pilot selection is given by
\begin{eqnarray}
    t_k &=& \mathop{\arg\min}\limits_{i | \zeta_{\ell,k,{i}} = 0 } \sum_{j \in \Pc_i} \beta_{\ell,j},
    \label{eq:pilot_choice_SP}
\end{eqnarray}
where $t_k$ is the least interfered pilot.
If the RU $\ell$ cannot satisfy the above conditions, the UE approaches the next candidate RU, or UEs are declared in outage if no RU is available.

\subsubsection{Cluster formation}
In the cluster formation of UE $k$, the cluster $\Cc_k$ is obtained sorting the RUs satisfying condition (\ref{eq:snr_threshold}) and $\zeta_{\ell,k,t_{k}} = 0 $ in decreasing LSFC order, and adding them to the cluster until a maximum cluster size $Q$ is reached.

\subsection{Rough overloaded pilot assignment}
\label{sec:overloadedPA}

\subsubsection{Cluster formation}
Finally, we introduce R-OPA, where user-centric clusters are formed before pilot allocation.
In this method, each UE $k$ chooses up to $Q$ RUs with condition $\beta_{\ell,k}>\frac{\eta}{M{\rm SNR} }$ in decreasing LSFC order.

\subsubsection{Pilot selection}
The pilots are assigned to each UE by random pilot assignment (RPA) or some strategic PA methods.
Here we propose a strategic PA based on a weighted graphic framework (WGF) based PA proposed in \cite{Zeng_WGF_2021}.
At first, we introduce WGF-based heuristic PA from \cite{Zeng_WGF_2021}, where subspace orthogonality is not considered.
The WGF scheme consists of two main phases: the construction of a weighted pilot contamination graph and Max k-Cut PA.
The aim of the Max k-Cut algorithm is finding the optimal $\tau_p$ co-pilot UEs sets $\{\Vc_1,\Vc_2,\ldots,\Vc_{\tau_p} \}$ so that the potential contamination is minimum.
However, implementation of the pure Max k-Cut algorithm has high complexity, thus \cite{Zeng_WGF_2021} introduces a heuristic approximation of the Max k-Cut.
The heuristic approach consists of the following steps, where each variable is explained later.
\begin{enumerate}
	\item Assign $\tau_p$ arbitrarily chosen UEs to $\tau_p$ subsets, one in each subset. Temporal subsets are given as $\Vc_1 = \{ \text{UE}_1 \}, \ldots, \Vc_{\tau_p} = \{ \text{UE}_{\tau_p} \} $.
	\item Select one remaining $\text{UE}_i$, calculate a weight between each subset and $\text{UE}_i$ as $W_{i,q} = \sum_{j \in \Vc_q} \omega_{i,j}$, where $\omega_{i,j}$ denotes a weight between two UEs.
    \item Assign the UE to the subset with the smallest increased weight as $q^* = \mathop{\arg\min}\limits_{q} W_{i,q}$ and update subset as $\Vc_{q^*}  = \Vc_{q^*}\cup \text{UE}_i$.
	\item Iteratively repeat step 2) and 3) until the remaining UEs are assigned.
\end{enumerate}
Firstly, a potential pilot contamination $\omega_{k,k'}$ between the $k$-th and $k'$-th UE is defined as
\begin{eqnarray}
    \omega_{k,k'} = 
        \left| \frac{ \sum_{\ell \in \Cc_k} \beta_{\ell, k'}}
        {\sum_{\ell \in \Cc_k} \beta_{\ell, k}} 
        \right|^2 
        + \left| \frac{\sum_{\ell \in \Cc_{k'}} \beta_{\ell, k} } 
        {\sum_{\ell \in \Cc_{k'}} \beta_{\ell, k'} } \right|^2 ,
    \label{eq:potential_interference}
\end{eqnarray}
where the first term on the right-hand side corresponds to the amount of interference from the $k'$-th UE to the $k$-th cluster $\Cc_k$, and the second term corresponds to interference from the $k$-th UE to $\Cc_{k'}$.
Secondly, a weight between subset $\Vc_p$ and $\Vc_q$ is defined as
\begin{equation} 
W_{p,q} = \sum_{i \in \Vc_p, j \in \Vc_q} \omega_{i,j}.
\label{eq:sum_potential_interference}
\end{equation}
This scheme tends to assign UEs with high potential interference  into the different subsets, thus UEs with potentially high interference are given orthogonal pilots.

For further enhancing the robustness to the contamination, we modify the WGF PA so that the potential interference after the SP becomes small, by taking the subspace information into account in the WGF metric.
%
Using SP with $\Fm_{\ell,k}\Fm_{\ell,k}^\herm$, the new metric is given by
\begin{eqnarray}
    \omega_{k,k'}^{\rm SP} &=& 
        \left| \sum_{\ell \in \Cc_k} 
        \frac{|| \Fm_{\ell,k}\Fm_{\ell,k}^\herm\Fm_{\ell,k'} ||_F^2 }{|\Sc_{\ell,k}|} \beta_{\ell, k'} \middle/ \sum_{\ell \in \Cc_k} \beta_{\ell, k}
        \right|^2 \nonumber \\
        &+& \left| \sum_{\ell \in \Cc_{k'}} 
        \frac{|| \Fm_{\ell,k'}\Fm_{\ell,k'}^\herm\Fm_{\ell,k} ||_F^2 }{|\Sc_{\ell,k'}|} \beta_{\ell, k} \middle/ \sum_{\ell \in \Cc_{k'}} \beta_{\ell, k'} \right|^2 
        .
    \label{eq:potential_interference_2}
\end{eqnarray}
Note that $|| \Fm_{\ell,k}\Fm_{\ell,k}^\herm\Fm_{\ell,k'} ||_F^2$ is given by $|\Sc_{\ell,k} \cap \Sc_{\ell,k'}|$.
If the subspaces of UE $k$ and $k'$ are orthogonal, i.e. $|\Sc_{\ell,k} \cap \Sc_{\ell,k'}|=0$, the potential interference after SP is 0 and $\omega_{k,k'}^{\rm SP}$ reflects this.
%
One point to consider in using (\ref{eq:potential_interference_2}) is that these expressions make use of the LSFCs of not associated RU-UE pairs, i.e., $\beta_{\ell, k^{\prime}}$ for pairs $(\ell,k^{\prime})$ such that $\ell \notin \Cc_{k^{\prime}}$. 
Since RU $\ell$ is not part of the cluster serving user $k^{\prime}$, such LSFCs may be difficult to be estimated and may not be available. On the other hand, if RU $\ell$ is not part of $\Cc_{k^{\prime}}$ it is likely that $\beta_{\ell,k^{\prime}}$ is very small (otherwise, the RU would likely be part of the cluster). 
Motivated by this, we introduce a partial LSFC-based WGF metric where only the LSFCs of associated UE-RU pairs are available, i.e.,
%
%
\begin{eqnarray}
    \overline{\omega}_{k,k'}^{\rm SP} &=& 
        \left| \sum_{\ell \in \Cc_k \cap \Cc_{k'}} \frac{|\Sc_{\ell,k} \cap \Sc_{\ell,k'}|}{ |\Sc_{\ell,k}|}  \beta_{\ell, k'} \middle/ \sum_{\ell \in \Cc_k} \beta_{\ell, k}
        \right|^2 \nonumber \\
        &+& \left| \sum_{\ell \in \Cc_{k'} \cap \Cc_k} \frac{|\Sc_{\ell,k} \cap \Sc_{\ell,k'}|}{ |\Sc_{\ell,k}|}  \beta_{\ell, k} \middle/ \sum_{\ell \in \Cc_{k'}} \beta_{\ell, k'} \right|^2 .
    \label{eq:potential_interference_3}
\end{eqnarray}
We use this partial LSFC-based metric in (\ref{eq:sum_potential_interference}) for numerical simulations.
The complexity (number of weights calculations and set operations) of the heuristic WGF method is given by $\mathcal{O}(K^2/2+K/2+\tau_p)$ \cite{Zeng_WGF_2021}, where the actual operational complexity depends on the hardware implementation.
On the other hand, non-overloaded PA and SIA-OPA are employed in a decentralized manner, where each RU mainly checks (\ref{eq:pilot_choice}) or (\ref{eq:pilot_choice_SP}) for each pilot and each UE, which are much less computationally complex than the WGF methods and do not require network-wide information exchange.


\section{Simulation results}
Table \ref{tab:sim_param} shows the basic parameter specifications of the simulations. 
\begin{table}[tb]
    \centering
    \caption{Simulation parameters}
    \begin{tabular}{|p{4cm}|p{4cm}|}
        \hline
        Area size ($A$) & $2\times 2 \ {\rm km}^2$  \\ \hline
        Number of UEs ($K$) & 100--1200 \\ \hline
        Number of RUs and RU antennas ($\{L,M\}$)  &
        $\{10,40\}, \{25,16\}$, $\{50,8\}, \{100,4\}$  \\ \hline
        Pilot dimension ($\tau_p$) & $10, 20, 30$ \\ \hline
        Maximum cluster size ($Q$) & 10 \\ \hline
        Cluster formation threshold ($\eta$) & 1 \\ \hline
        Pathloss model & 3GPP urban microcell channel \cite{3gpp38901} with 3.7 GHz carrier frequency\\ \hline
        Bandwidth per UE & 10 MHz \\\hline
        Transmit power per UE & 20 dBm \\ \hline
        Noise power spectral density & $-174$ dBm/Hz \\ \hline
        Dimension of a RB ($T$) & 200\\\hline
        Angular spread ($\Delta$) & $\pi/8$ \\ \hline
    \end{tabular}
    \label{tab:sim_param}
\end{table}
We consider a square coverage area of $A= 2 \times 2 \ {\rm km}^2$ with a torus topology by using the wrapped around technique to avoid boundary effects.
LSFCs are given according to the 3GPP urban microcell street canyon pathloss model from \cite[Table 7.4.1-1]{3gpp38901}, which differentiates between UEs in LOS and NLOS.
The total number of receiving antennas is fixed as $LM = 400$.
For each set of parameters we generate 40 independent layouts, and small scale fading coefficients are varied 30 times for each layout.
RUs and UEs are randomly distributed in each setup, and the sum SE is given by the sum of ${\rm SE}_k^{\rm ul}$ over all $k \in [K]$, and then averaging it with respect to the different layouts.

\begin{table}[tb]
    \centering
    \caption{The largest sum SE with parameters}
    \begin{tabular}{|c|c|c|c|c|c|}
        \hline
        PA scheme &  
        \begin{tabular}{c}
        Sum SE \\ \ [bit/s/Hz]
        \end{tabular}
        & $K$ & $L$ & $M$ & $\tau_p$\\ \Hline
        \begin{tabular}{c}
        Non-overloaded PA \\ with PM  
        \end{tabular} & 738 & 800 & 100 & 4 & 20\\ \hline
        \begin{tabular}{c}
        Non-overloaded PA \\ with SP  
        \end{tabular} & 770 & 800 & 100 & 4 & 20\\ \hline
        \begin{tabular}{c}
        SIA-OPA with SP  
        \end{tabular} & 866 & 600 & 25 & 16 & 30\\ \hline
        \begin{tabular}{c}
        R-OPA Random with SP  
        \end{tabular} & 740 & 400 & 50 & 8 & 30\\ \hline
        \begin{tabular}{c}
        R-OPA WGF {with SP}  
        \end{tabular} & 841 & 600 & 25 & 16 & 30\\ \hline
        \begin{tabular}{c}
        R-OPA WGF  {with PM}  
        \end{tabular} & 534 & 200 & 50 & 8 & 30\\ \hline
    \end{tabular}
    \label{tab:largest_sum_SE}
\end{table}
Table \ref{tab:largest_sum_SE} 
shows the largest sum SE achieved by the different PA schemes and the corresponding system parameters, where R-OPA Random denotes rough overloaded PA using random pilot assignment and R-OPA WGF denotes rough overloaded PA using WGF-based pilot assignment.
Firstly, we confirm that $\tau_p = 30$ outperforms $\tau_p = 20$ for the R-OPA and the SIA-OPA even taking into account the overhead in eq. (\ref{eq:SE}), while $\tau_p = 20$ is the best for the non-overloaded PA. 
As a comparison with existing studies, the results of R-OPA with the WGF algorithm and PM channel estimation represent \cite{Zeng_WGF_2021}.
From Table \ref{tab:largest_sum_SE}, the SIA-OPA with SP achieves the highest sum SE.
In the following, we focus on the setup $\tau_p = 30, L = 25, M = 16$ to analyze the behavior around the largest sum SE.

\begin{figure}[t]
    \centering
    \includegraphics[keepaspectratio, scale=0.57]{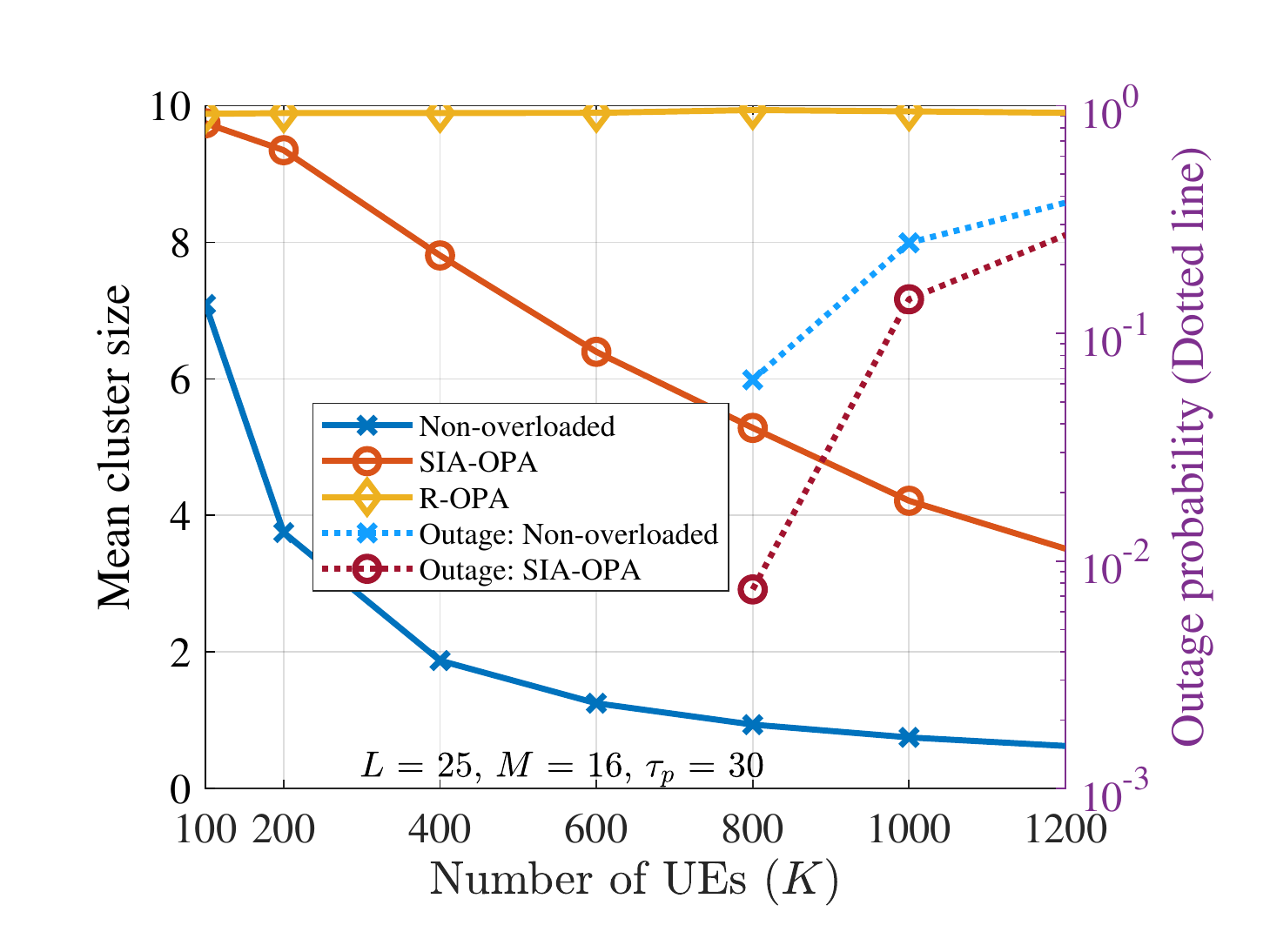}
    \caption{Mean cluster size and outage probability vs. $K$.}
    \label{fig:cluster_size}
\end{figure}
\begin{figure}[t]
    \centering
    \includegraphics[keepaspectratio, scale=0.57]{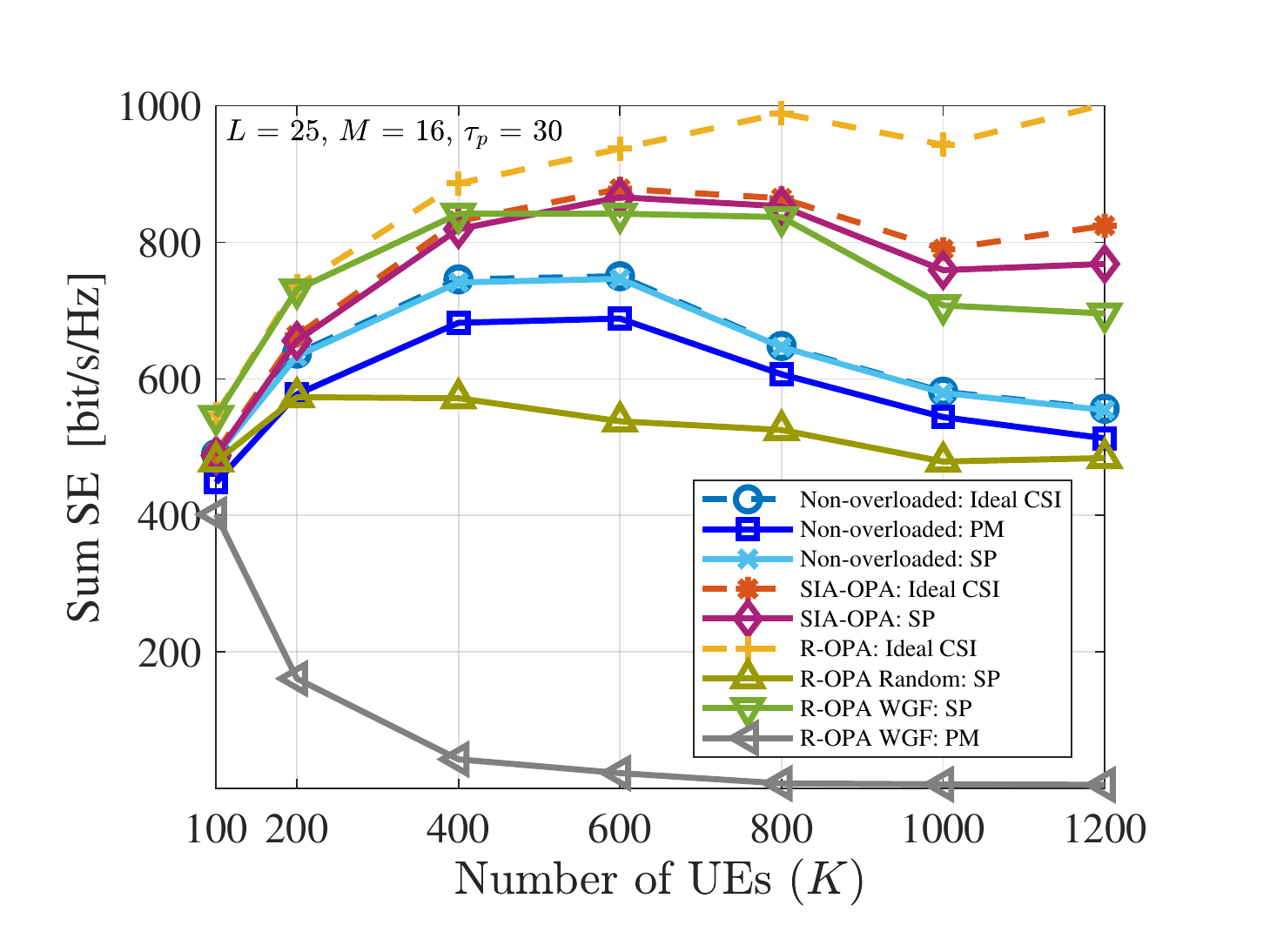}
    \caption{Sum SE vs. $K$.}
    \label{fig:sumSE_vs_K}
\end{figure}

Fig. \ref{fig:cluster_size} shows the mean cluster size and outage probability vs. $K$, and Fig. \ref{fig:sumSE_vs_K} illustrates the sum SE vs. $K$.
Note that outage occurs in the region of more than 800 UEs for the SIA-OPA and the non-overloaded PA.
From Fig. \ref{fig:cluster_size}, the cluster size with rough overloaded PA (R-OPA) is almost equal to $Q = 10$ for all evaluated $K$.
From this observation we deduce for the given simulation parameters that if there is no limitation on cluster formation imposed by the pilot assignment such as non-overloaded, $Q$ RUs with the largest channel gain will form a user-centric cluster for most UEs.
Note that the cluster size with R-OPA is independent of the pilot assignment scheme, since the clusters are formed before assigning pilots.
The cluster size of the non-overloaded PA scheme is significantly reduced by the increase of $K$, while that of the SIA-OPA scheme decreases more gradual.
For small channel estimation error, UEs with larger clusters can be given higher SE by obtaining more spatial diversity.
Thus, the cluster formation of R-OPA and SIA-OPA have potential to outperform the non-overloaded PA in terms of SE.

Then let us focus on Fig. \ref{fig:sumSE_vs_K} where Ideal CSI lines indicate no channel estimation error.
The SE degradation factor can be divided into two: channel estimation error due to pilot contamination, 
and deviation from the optimal multiplexing gain of massive MIMO.
The fact that the sum SE saturates in the ideal CSI case indicates the latter factor.
%
%

The R-OPA WGF scheme with PM channel estimation is not tolerant of an increase in UE density. This is an interesting effect when compared to the non-overloaded PA with PM, where the sum SE first grows with the number of UEs, and then decreases again. The degradation of the R-OPA WGF scheme with PM is mainly due to an increased channel estimation error, while the non-overloaded PM is specifically designed to avoid assigning the same pilot to UEs that can potentially cause large mutual pilot contamination.
The R-OPA WGF scheme allows co-pilot UEs with large potential interference if their subspaces are mutually orthogonal. Since the PM channel estimation only eliminates pilot contamination from UEs with a different UL pilot (and does not take into account the channel subspaces), the channel estimate is potentially contaminated to a large extent by co-pilot users with orthogonal subspaces, which leads to a strong degradation of the sum SE for growing $K$.

The non-overloaded PA with the SP achieves almost the same sum SE as the ideal case, which means that the SP effectively eliminates the contamination.
For the R-OPA, the performance of random assignment significantly degrades compared to ideal CSI as $K$ increases due to that randomly assigned pilots generate many co-pilot UEs within the same subspace.
The WGF-based R-OPA also degrades compared to the ideal case, however, it achieves the highest sum SE at $K \leq 400$ among all PA methods.
At last, SIA-OPA slightly outperforms the R-OPA at $K \geq 600$ and achieves the highest sum SE at $K = 600$.
In addition, the gap between the SP and the ideal CSI stays very small, thus the SIA-OPA works as expected.
In the region of $K \geq 800$, the sum SE saturates and slightly degrades, which can be considered the effect of increased outage as seen in Fig. \ref{fig:cluster_size}.

The WGF R-OPA has the highest sum SE at a certain UE density ($K \leq 400$), where the per UE SE (sum SE$/K$) is the largest.
On the other hand, the SIA-OPA achieves a competitive sum SE at $K = 400$ and is easier to implement in terms of scalability than WGF R-OPA.


\section{Conclusions}

This paper has investigated overloaded PA and cluster formation methods that outperform the non-overloaded PA in CF-mMIMO systems.
The WGF R-OPA achieved the highest sum SE in the region of a certain UE density ($K \leq 400$), while it is challenging for practical implementation due to the unscalable metric calculation and information exchange.
The SIA-OPA, which assigns pilots to UEs with orthogonal subspaces, achieves the highest sum SE in the region of $K \geq 600$.
Furthermore, it also achieves a competitive sum SE at $K \geq 400$ with a more relaxed PA procedure.
The appropriate method in a practical system can be chosen according to the trade-off between their performance and feasibility.

\bibliography{IEEEabrv,CF_PA.bib}

\end{document}